\documentclass[jphysg,11pt]{article}
\usepackage[dvips]{graphics}
\usepackage{epsfig}

\def\lapproxeq{\lower .7ex\hbox{$\;\stackrel{\textstyle<}{\sim}\;$}}
\def\gapproxeq{\lower .7ex\hbox{$\;\stackrel{\textstyle>}{\sim}\;$}}

\title{The Anisotropy of Galactic Cosmic Rays as a Product of Stochastic Supernova 
Explosions}

\author{A. D. Erlykin$^{1,2}$ and A. W. Wolfendale$^2$\\[1ex] $^1$PN Lebedev
Physical Institute, Moscow, Russia\\ $^2$Department of Physics,
University of Durham, Durham, UK}

\date{\today}

\begin{document}

\def\gtrsim{ \;\raisebox{-.7ex}{$\stackrel{\textstyle
>}{\sim}$}\; }
\def\lesim{ \;\raisebox{-.7ex}{$\stackrel{\textstyle
<}{\sim}$}\; }

\maketitle

\begin{abstract}
We study the effect of the stochastic character of supernova explosions on the 
anisotropy of galactic cosmic rays below the knee. We conclude that if the bulk of 
cosmic rays are produced in supernova explosions the observed small and nearly energy 
independent amplitude of the anisotropy and its phase are to the large extent 
determined by the history of these explosions in the vicinity of the solar system, 
namely by the location and the age of the supernova remnants, within a few kpc, which 
give the highest contibution to the total 
intensity at the present epoch. Among the most important factors which result in the
small magnitude and the energy independence of the anisotropy amplitude are the mixed 
primary mass composition, the effect of the Single Source and the Galactic Halo. 
Special attention is given to the phase of the anisotropy. It is shown that the 
excessive cosmic ray flux from the Outer Galaxy can be due to the location of the Solar
 System at the inner edge of the Orion Arm which has the enhanced density and rate of 
supernova explosions. 
\end{abstract}

\section{Introduction}

Since the discovery of cosmic rays (~CR~) the study of their arrival directions is one 
of the most popular studies aiming eventually to discover their origin. The 
difficulty of this study is an extremely uniform distribution of CR arrival 
directions. The anisotropy amplitude in the energy range from tens of GeV to sub-PeV 
is less than 10$^{-3}$ and does not show any significant energy dependence (~see the 
recent summary in \cite{Guil} and references therein~). It is surprising since the 
amplitude $A$ is usually connected with the diffusion coefficient $D$ as
$A = \frac{3DgradI}{cI}$, where $I$ is the CR intensity and $c$ is 
the speed of light. Since the diffusion coefficient is admitted to rise with the 
energy \cite{Gais} as $E^s$ with $s \approx 0.3-0.6$, then one would expect a 
corresponding rise of the anisotropy, which in fact is not observed \cite{Hill}. 

The view of the majority of CR workers about CR origin is that at energies below a PeV 
CR are produced mainly by the shocks from supernova (~SN~) explosions. These explosions
 are a stochastic process and the importance of the fluctuations in the cosmic ray 
production and propagation has been underlined in many works 
\cite{Jones,Lee,Berez,Pohl}. The dominant role in these fluctuations is played by
explosions of nearby and recent SN \cite{Ahar,Nish,EW1,EW2}. The anomalous diffusion 
in the non-uniform interstellar medium (~ISM~) magnifies these fluctuations compared 
with normal 'gaussian' diffusion \cite{Lag1,EW3}. Attempts to explain the observed 
anisotropy by the random character of SN explosions have been made in \cite{Ptus,Svesh}
 and by the anomalous diffusion - in \cite{Lag2}. In the present paper we analyse 
further the anisotropy of CR in the framework of scenario with the stochastic nature 
of SN explosions and the subsequent anomalous diffusion through the ISM of our Galaxy. 
Special attention is devoted to the observed phase of the anisotropy.

\section{Simulations}

We simulated 10$^6$ random explosions of SN in our Galaxy uniformly distributed in the 
time interval of 10$^8$ years, so that the explosion rate of 10$^{-2} year^{-1}$ 
corresponds to the explosions of Type II SN, which give the dominant contribution to 
the CR intensity. The spatial distribution of SN in the Galactic Disk had cylidrical 
symmetry with the surface density \cite{Case}:
\begin{equation}
\rho_{SN} \propto (\frac{R}{R_\odot})^{1.69}exp(-3.33\frac{R}{R_\odot}) 
\end{equation}
and a vertical distribution:
\begin{equation}
\frac{d\rho_{SN}}{dZ} \propto exp(-\frac{|Z|}{200})
\end{equation}
Here $R$ and $Z$ are the Galactocentric radius and the height above or below the 
Galactic Plane in {\em pc}, $R_\odot = 8500 pc$ is the Galactocentric radius of the 
Sun. The energy of each SN explosion has been taken fixed and equal to 10$^{51}$ erg,
from which 10$^{50}$ erg is transferred to CR. The model of the explosion and 
subsequent generation of shocks was taken 
as in \cite{EW1}. The maximum rigidity of the accelerated particles is 0.4 PV. 
The CR propagation in the ISM has been treated in the framework
 of the anomalous diffusion model with the intrinsic parameter $\alpha = 1$
\cite{Lag1,EW2}. We prefer this model since \\ 
(i) our ISM is certainly non-uniform and has quasi-fractal properties;\\
(ii) this model helps us to understand some of the observed features: the small radial 
gradient of CR in the Galaxy, the formation of the Galactic Halo, the Galactic Plane 
enhancement at the sub-PeV energy \cite{EW2} etc. \\
We simulated also the CR propagation using an ordinary 'Gaussian' diffusion 
model, but it gave basically the same results, so that we present here results only 
for the anomalous diffusion model. The diffusion coefficient $D$ in our 
model increases with energy as $D \propto E^{\alpha/2}$, i.e. for $\alpha = 1$ 
it rises as $\sim \sqrt{E}$. In the present scenario no effects due to
the regular galactic magnetic field have been considered and those due to irregular
magnetic fields were incorporated into the general characteristics of the diffusion 
process. The diffusion has been assumed to be spherically symmetric.

Within the framework of the anomalous diffusion model the anistropy amplitude has been
 calculated using the CR flux vector averaged over the entire ensemble of 10$^6$ SN.   
The absolute value of the flux from a single SN has been calculated for the case of 
the spherically symmetric diffusion as 
\begin{equation}
{\bf |j_i|} = grad(\frac{dI_i}{dt}) = \frac{1}{r_i^2}\int_{r_i}^{\infty} \frac{dI_i}{dt}(l,t)l^2 dl
\end{equation}
where ${\bf |j_i|}$ is the absolute value of the flux vector, $I_i$ is the CR intensity
, $l$ is a variable distance from the SN, $r_i$ is the distance from the SN to the Sun 
and $t$ is the age of the SN \cite{Lag2}. 
The direction of the flux vector was taken from the SN toward the Sun, so that the  
components of this vector were taken as $j_{x,y,z}^i={\bf |j_i|}cos\theta_{x,y,z}^i$ 
where $cos\theta_{x,y,z}^i$ are cosines of the angles between the flux vector and 
coordinate axes. 

The aggregate vector components for the entire ensemble of SN have been obtained by
 summing $j_{x,y,x}^i$ from all 10$^6$ SN:
\begin{equation}
J_{x,y,z} = \sum_i j_{x,y,z}^i
\end{equation}
and the absolute value of the total flux vector has been calculated as
\begin{equation}
{\bf |J|} = \sqrt{J_x^2 + J_y^2 +J_z^2}
\end{equation}
Similarly, the total CR intensity $I$ has been obtained as the sum of the contributions
 $I_i$ from all 10$^6$ SN: 
\begin{equation}
I = \sum_i I_i
\end{equation}
The anistropy amplitude has been obtained as
\begin{equation}
A = \frac{3{\bf |J|}}{cI}
\end{equation}
The galactic phase of the anisotropy has been taken in the direction opposite to the 
projection of the flux vector on the Galactic Plane, i.e. as the galactic longitude 
{\em from} which the maximum total flux ${\bf |J|}$ is observed (~the Y coordinate axis
 corresponds to the direction from the Galactic Centre to the Sun~):
\begin{equation}
L_{max} = arctan(\frac{J_x}{J_y})
\end{equation}

\section{Results} 
\subsection{Primary protons}

In Figure 1 we show the simulation results for 50 samples of supernova remnants (~SNR~)
 emitting only primary protons. Figure 1a shows the energy spectra of these protons. 
As in our previous publications one can see that although there is a general trend to 
create the power law spectrum with a slope index about 2.65 the fluctuations about this
mean trend are large. The same evidence is observed for the anisotropy amplitude 
shown in Figure 1b. The mean amplitude rises with energy as $\sim E^{0.5}$, which
corresponds to the adopted energy dependence of the diffusion coefficient, but the 
fluctuations about this mean behaviour are quite large. In some samples one can even 
get a decrease of the amplitude instead of an increase - particularly at higher 
energies where the effect of the nearby and recent SN is greatest.  The phases of 
the anistropy (~Figure 1c~) are mostly concentrated around the direction to the 
Galactic Centre (~$L_{max} = 0^\circ$, mind that here the phase is in galactic and not 
in equatorial coordinates~), but demonstrate the large variations around 
this direction for different energies even within the same sample of SN. Some of the 
samples have even opposite phases, i.e. the resultant flux {\em at some energies} comes
 from the Galactic Anticentre (~$L_{max} \approx 180^\circ$~).
\begin{figure}[htb!]
\begin{center}
\includegraphics[height=15cm,width=15cm]{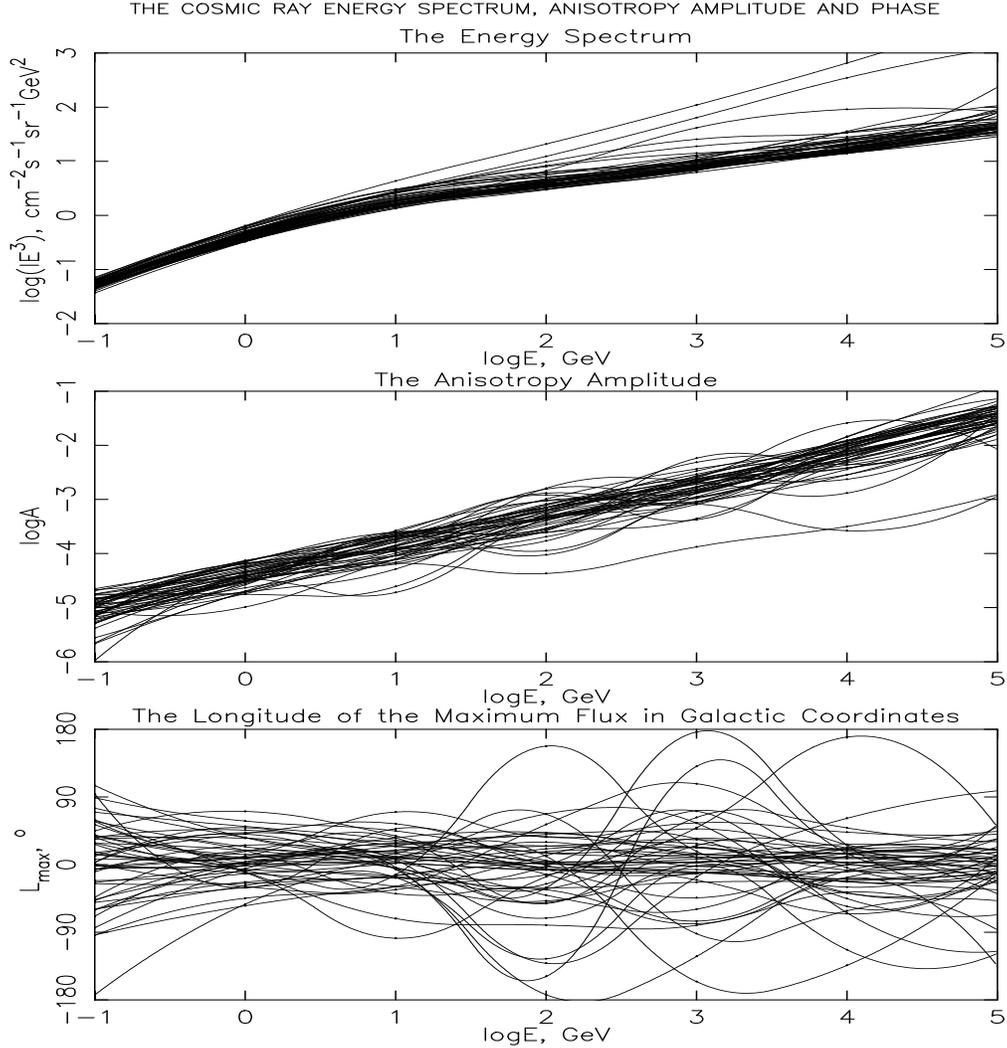}
\caption{\footnotesize Primary protons in 50 samples of 10$^6$ SN which exploded 
sporadically in our Galaxy.
(a) Energy spectra of protons. (b) The anisotropy amplitude as the function of energy.
(c) The galactic phase of the anisotropy as a function of energy.}
\end{center}
\label{fig:anis1}
\end{figure}    

Figure 2 shows the correlation between the amplitude and the galactic phase at 4 
primary 
energies: 10$^2$, 10$^3$, 10$^4$ and 10$^5$ Gev for 100 different samples of SN.
The concentration around $L_{max} = 0^\circ$, the general rise of the amplitude with 
the energy and large fluctuations between different samples of SN are clearly seen. 
These fluctuations are the consequence of the stochastic character and the dominant 
role of recent and local SN explosions and their subsequent SNR.     
\begin{figure}[htb!]
\begin{center}
\includegraphics[height=15cm,width=12cm,angle=-90]{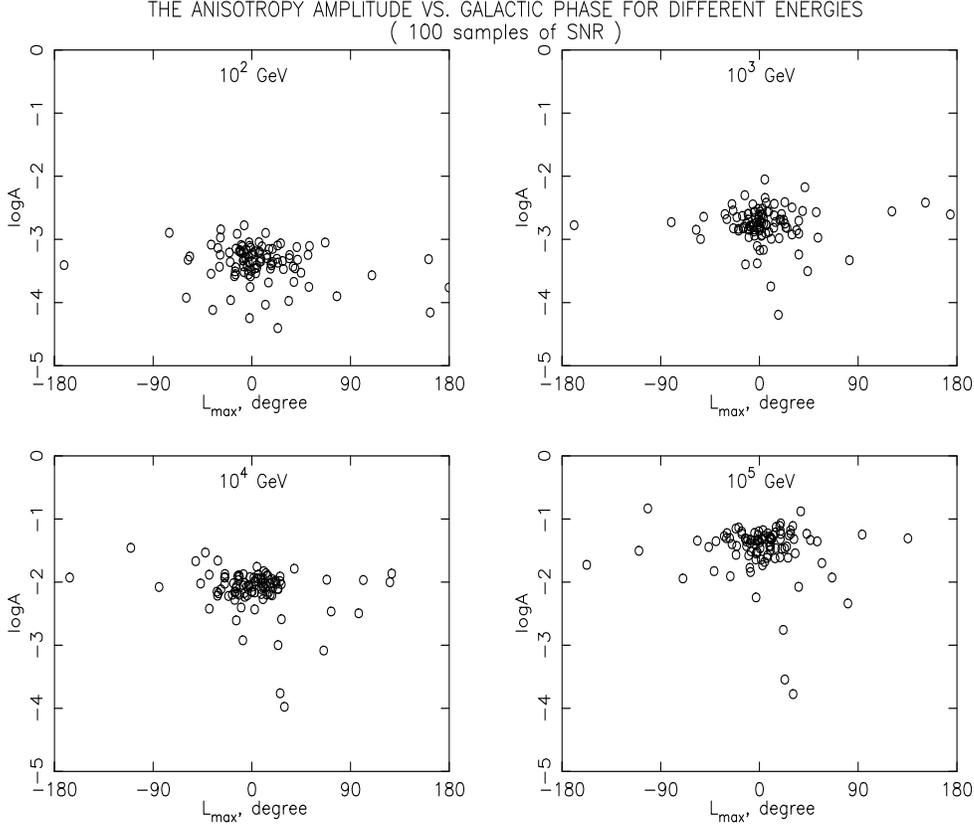}
\caption{\footnotesize The anisotropy amplitude vs. the galactic phase at different 
energies for 100 simulated samples of SN which emit only protons: (a) 10$^2$ Gev, 
(b) 10$^3$ GeV, (c) 10$^4$ GeV, (d) 10$^5$ GeV.}
\end{center}
\label{fig:anis2}
\end{figure}    

\subsection{Mixed primary mass composition}

One of the factors leading to the observed small anisotropy is the mixed CR mass 
composition. In our model, both the CR acceleration and the propagation depend on the 
particle rigidity and in this case the maximum flux {\em for the different 
primary nuclei at the given energy} can be from different supernova remnants. As has
been described in \S2 the total CR flux is obtained as the summation of partial 
flux {\em vectors}. For the mixed primary mass composition the destructive interference
 of fluxes from different nuclei coming from different directions results as a rule in 
a considerable loss of the directionality and therefore the reduction of the 
anisotropy amplitude. 
In Figure 3 we show one of the SN samples in which SN produce 
not only protons (~P~), but also helium (~He~), oxygen (~O~) and iron (~Fe~) nuclei.
In Figure 3a the energy spectra of these nuclei are shown taking each nucleus in turn,
viz. all the CR energy (~10$^{50}$erg~) going into P, or He, or O, or Fe. It is seen 
that the maximum energy for each nucleus is shifted to higher values by the factor 
equal to its charge $Z$. The kinematic effect due to the higher mass results also in 
some shift of the absolute intensity for heavier nuclei. 
\begin{figure}[htb!]
\begin{center}
\includegraphics[height=15cm,width=15cm]{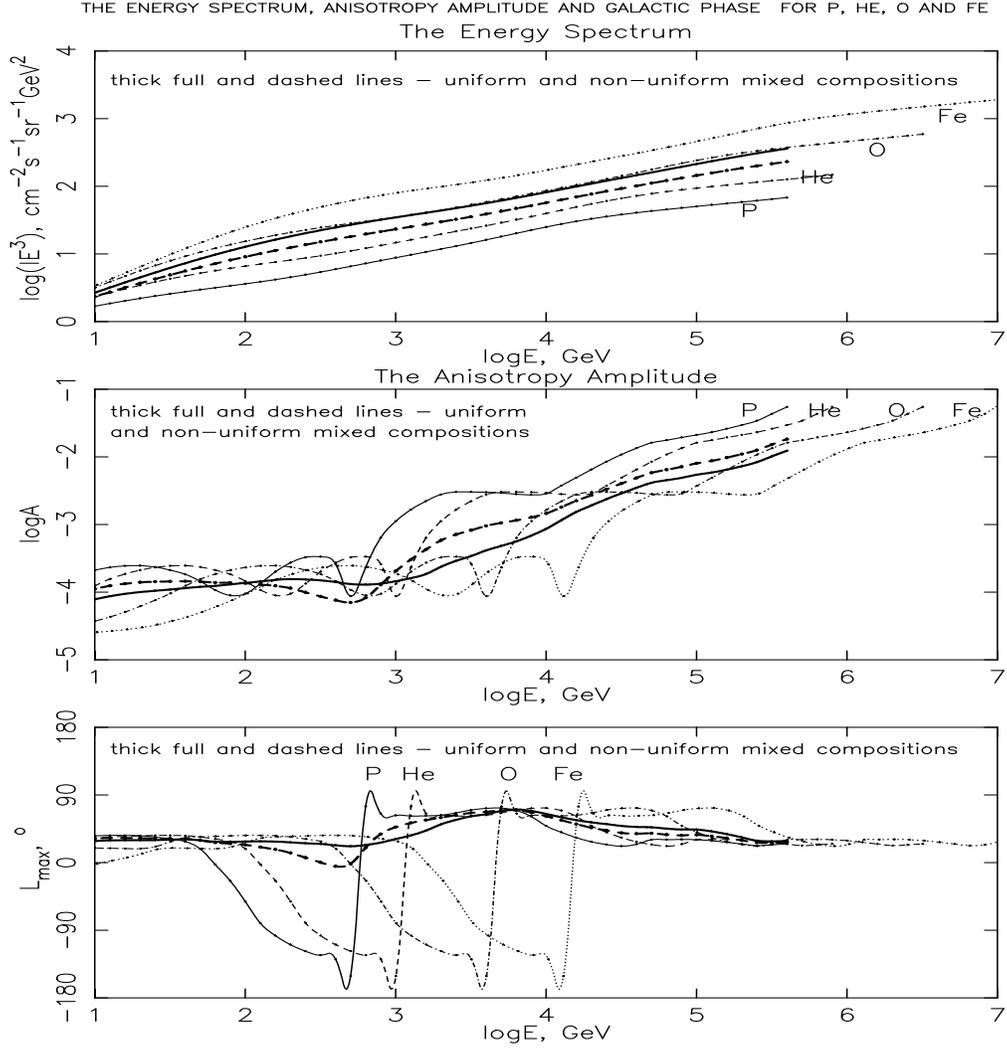}
\caption{\footnotesize An example of the energy spectra, the anisotropy amplitudes and 
galactic phases for alternatives where the primaries are protons (~P~), helium (~He~), 
oxygen (~O~) or iron (~Fe~) nuclei and for the mixed mass composition. (a) The energy 
spectra, (b) the anisotropy amplitudes 
and (c) the galactic phases. The thick full lines in all figures show the relevant 
mean characteristics for the uniform mixed primary mass composition with the abundance 
for each nucleus equal to 0.25. The thick dashed lines are for a non-uniform mass 
composition with the dominance of protons and helium nuclei: 0.43-P, 0.27-He, 0.19-O,
0.11-Fe. The destructive interference of the phases leads to the
 nearly energy independent total anisotropy amplitude at 10 - 1000 GeV inspite that all
 partial amplitudes show strong variations in the same energy range.}
\end{center}
\label{fig:anis3}
\end{figure}    
Figures 3b and 3c show the corresponding anisotropy amplitude and the galactic phase 
for each nucleus. They also look as if they were obtained by the shift of the 
corresponding 
curve for protons by the same Z factor. However due to this shift and for some 
particular space-time configuration of SN the positive flux for one nucleus can be 
partly compensated by the negative flux of 
another nucleus and the resulting anisotropy amplitude for the mixed primary mass 
composition might be less than amplitudes of the constituent nuclei. 

Calculations have also been made for the (~hypothetical~) situation where SNR 
accelerate all four types of nucleus - with equal energy contents - this last-mentioned
 situation is denoted {\em 'uniform mixed composition'}. In spite of the strong energy 
variations of all partial amplitudes the destructive interference of the phases leads 
to a total amplitude for the mixed mass composition which is nearly energy independent 
in the 10 - 1000 GeV energy range. The resulting energy spectrum, the anisotropy 
amplitude and the phase for the uniform mixed primary mass composition are all shown by
 thick full lines in Figures 3a, 3b and 3c.

In fact, equipartition of the different masses is not the situation in practice; the 
division being such that the fraction of 'heavy' nuclei increases with energy from a 
low base. The 'true' reduction in anisotropy caused by the addition of 'non-protons'
will thus be somewhat smaller than shown in Figure 3b. However, this is offset, 
somewhat, by variations in the mass spectrum from one SNR to another due to varying 
composition of the accelerated medium, such variations being due to variations in the 
ISM itself together with the (~variable~) material ejected from the SNR precursor star.
In Figure 3 we show by thick dashed lines an example of the non-uniform mass 
composition with the dominance of protons and helium nuclei: 0.43 - P, 0.27 - He, 
0.19 - O and 0.11 - Fe (~sub-Fe elements  are distributed between O and Fe groups~).
 
\subsection{The possible effect of the Single Source}          

Another possible factor which can compensate the expected rise of the anisotropy with 
energy is the Single Source (~SS~) - the nearby and recent supernova which 
{\em accidentally} exploded in the direction from the Sun downstream of the main CR 
flux. The flat energy spectrum of CR from this SS gives the rising 
contribution to the total CR intensity which becomes dominant at the knee 
\cite{EW4}. Due to the opposite phase of its flux this SS compensates the otherwise 
rising main flux and can reduce the total amplitude of the anisotropy at high 
energies. 

One of the likely candidates for the SS is the Monogem Ring SNR and associated pulsar 
B0656+14 \cite{EW5,EW6}. They are in the Outer Galaxy 
with Galactic coordinates $l \approx 200^\circ, b \approx 8^\circ$ at a distance of  
about 300 pc from the Sun. The age of the SNR is about 90 kyear and the energy of the 
explosion is $0.2\cdot10^{51}$ erg. We calculated the characteristics of CR from this 
SS assuming that it emits predominantly helium nuclei with a flat energy spectrum which
 eventually become dominant at the knee. Then we added them to CR shown in Figure 3 as 
an example. The result is shown in Figure 4.    
\begin{figure}[htb!]
\begin{center}
\includegraphics[height=15cm,width=15cm]{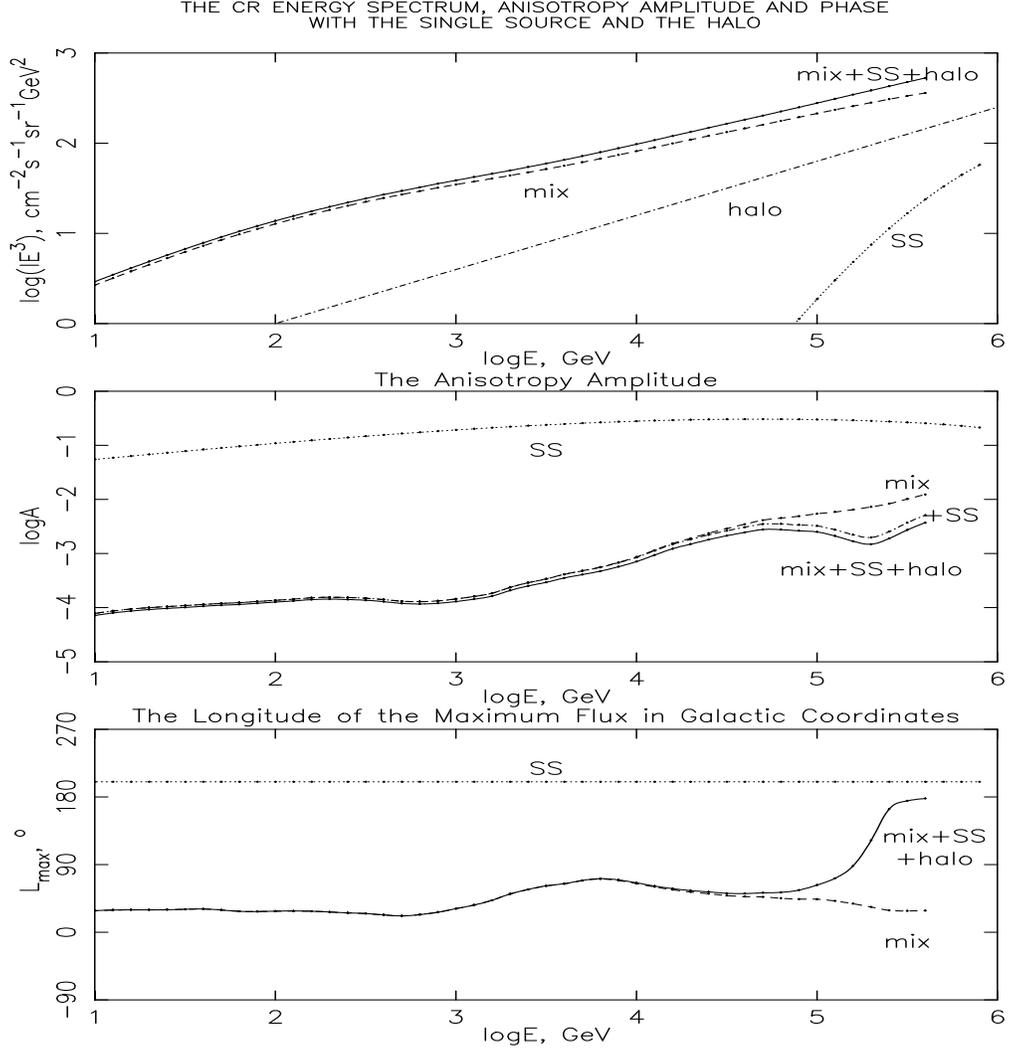}
\caption{\footnotesize The formation of the CR energy spectrum, the anisotropy 
amplitude and galactic phase for the scenario including the explosion of a nearby and
 recent SNR in the vicinity of the Sun called the Single Source (~SS~), and the 
Galactic Halo. Parameters of the SS were taken equal to those of the Monogem Ring SNR 
\cite{EW5}. Parameters of the Galactic Halo have been taken from \cite{EW7} with the 
slope index of the energy spectrum equal to 2.4. It is seen that 
the destructive interference of the flux from the Inner Galaxy and from the SS can 
lead to the reduction of the anisotropy when the contribution from the SS becomes 
substantial. A further reduction is caused by the Galactic Halo.}  
\end{center}
\label{fig:anis4}
\end{figure}    

It is seen that the SS flux streaming from the Outer Galaxy can compensate to some 
extent the main flux from the Inner Galaxy in the energy region above 10$^5$ GeV, 
where the contribution from the SS becomes substantial. This effect results in a 
reduction of the anisotropy amplitude. At even higher energies, where the flux from the
 SS overcomes that from the Inner Galaxy, one might expect a further rise of the 
amplitude and a change of the phase.
       
\subsection{The Galactic Halo}

One more factor which can reduce the anisotropy amplitude is the Galactic Halo.
In \cite{EW7} we advocated the existence of a Giant Galactic Halo as a mechanism needed
to smooth the irregularity of the CR energy spectrum produced by stochastic 
explosions of SN (~see Figure 1a~). CR accumulated in such Halo are completely 
isotropised and their contribution to CR from the Galactic Disk reduces the anisotropy.
For the illustration we show in Figure 4 the possible effect of the Galactic Halo. 
The parameters of the Halo have been taken from \cite{EW7}. The absolute intensity of 
CR in the Halo was normalized to the observed CR energy spectrum at 10$^8$ GeV and the 
slope index at lower energies was equal to 2.4. The reduction of the anisotropy 
amplitude due to this effect at sub-PeV energies is about 30\%-40\%. The phase of the 
anisotropy does not change with the addition of the isotropic component of the 
Galactic Halo.     

\section{Comparison with the experimental data}

A survey of experimental measurements of the amplitude and the phase of the first 
harmonic of the anisotropy is shown in Figures 5a and 5b. The phase is given in 
equatorial coordinates - the right ascension (~RA~). The figure has been reproduced 
from \cite{Guil} and includes the data from Super-Kamiokande I as well as the recent 
results from Tibet III \cite{Amen}, Baksan and Andyrchi \cite{Kozy}, final results 
from EAS-TOP \cite{Agli} and upper limits from KASCADE \cite{Anto}. Thin lines in 
Figure 5a shows the amplitude of 10 samples of SN including the Single Source and the 
Halo. These lines demonstrate the possibility for the SS and the Halo to reduce or even
 invert the rise of the anisotropy amplitude though the calculated values beyond 
$10^4$GeV are still above the experimental points. 
\begin{figure}[htb!]
\begin{center}
\includegraphics[height=15cm,width=15cm]{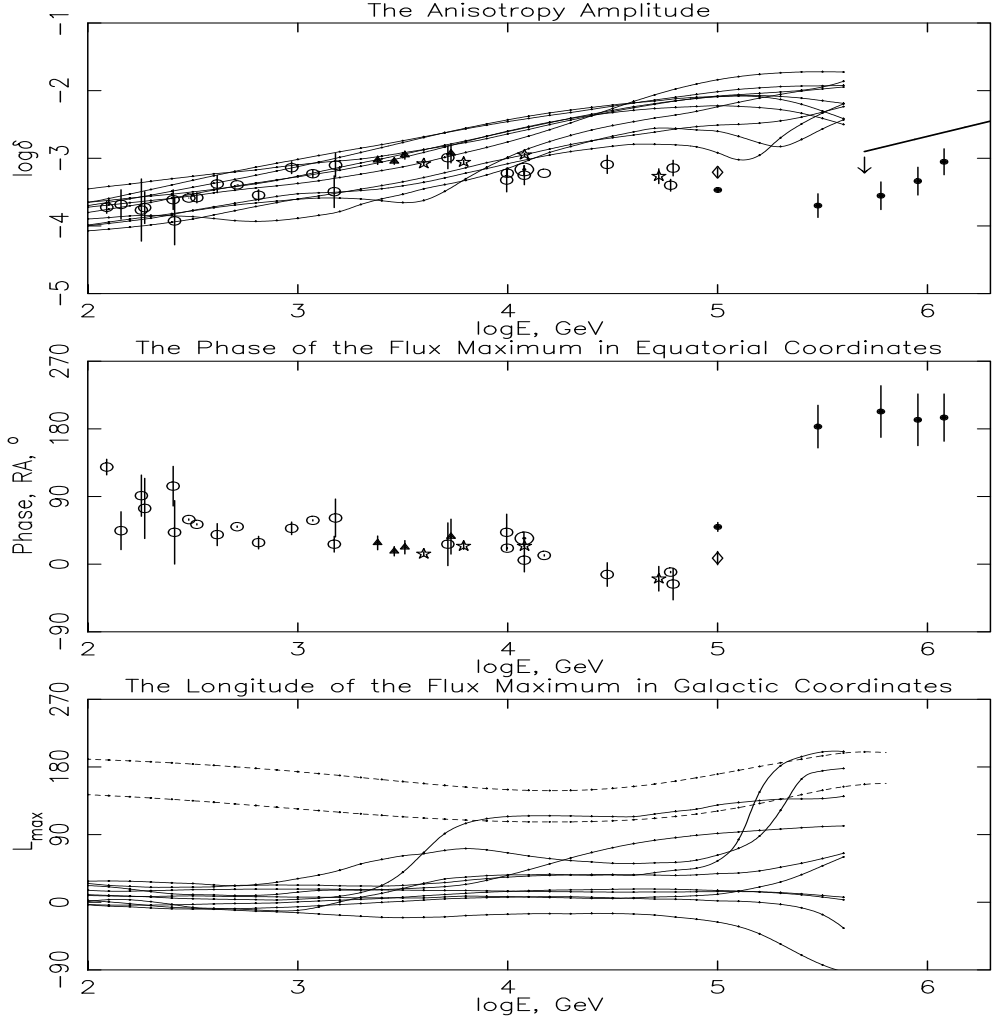}
\caption{\footnotesize The observed amplitude (a) and equatorial phase (b) of the 
first harmonic of the CR anisotropy. Data denoted as $\bigcirc$ are taken from the 
survey presented 
in \cite{Guil}, $\odot$ - Super-Kamiokande-I \cite{Guil}, $\star$ - Tibet III 
\cite{Amen}, $\triangle$ - Baksan \cite{Kozy}, $\diamond$ - Andyrchi \cite{Kozy},
$\bullet$ - EAS-TOP \cite{Agli}. Thick line above $logE = 5.7$ with an arrow - upper 
limits of the amplitude, given by KASCADE \cite{Anto}. Thin full lines - the 
calculations for 10 different configurations of SNR including a Single Source, which is
 tentatively associated with the Monogem Ring SNR. (c) The galactic phase of the 
anisotropy for the same 10 simulated samples of SN as shown in (a). The dashed lines
denote the region in galactic coordinates which corresponds to the RA region occupied
by experimental points in equatorial coordinates shown in (b) assuming that the 
observations were made at declinations about 30$^\circ$-60$^\circ$.}  
\end{center}
\label{fig:anis5}
\end{figure}    

The observed phases of the anisotropy are shown in Figure 5b. At the first sight they
indicate a change of the phase at an energy above 10$^5$GeV. A similar observation
on the possible change of the phase at the energy approaching 10$^6$GeV has been 
made in \cite{Clay}. It is remarkable, and probably significant, that the change in the phase occurs at the same 
sub-PeV region where there is an indication of the changing anisotropy amplitude 
(~Figure 5a~). 

It is tempting to connect these changes with the effect of the SS. 
However one has to be careful with such conclusion. First, the impression about these 
changes is based on the points of just one experiment: EAS-TOP \cite{Agli}, which 
certainly need an independent confirmation. Second, the closeness of the calculated 
amplitudes to the experimental points for some samples of SN does not mean that the 
described model which includes the SS and Halo can give an explanation of the gross 
features of the observed anisotropy. The measurements of the phase shown in Figure 5b 
are made in equatorial coordinates. If transferred to the galactic coordinates they 
occupy the region delimited approximately by two dashed lines in Figure 5c. It is seen 
that galactic phases calculated for 10 samples shown in Figure 5a have nothing common 
with this region. While calculated fluxes at energies $10^2 - 10^4$GeV are mainly from 
the Inner Galaxy the observed fluxes are predominantly from the Outer Galaxy 
particularly from its second quadrant. It means that other local phenomena play a 
role and should be also taken into consideration. These local phenomena will now be 
considered.

\section{The Phase of the Anisotropy and the Orion Arm}
The described model, in spite of the realistic inputs, such as the radial 
distribution of SN, the anomalous diffusion, the real existence of the Monogem Ring SNR
 etc. is nevertheless too simplified. It does not take into account other local 
phenomena: the geometry of the regular magnetic fields in the vicinity of the Sun, the 
real position of the Sun at the inner edge of the Orion arm with its enhanced rate of 
SN explosions etc. So far we have used the SNR radial distribution constructed
 and fitted by Case and Bhattacharya (~formula 1~) \cite{Case}. In fact their 
distribution contained 81 SNR found in the 60$^\circ$-wide sector of the Galactic Disk 
closest to the Sun. The fit describes only the general shape of the distribution and 
has a big uncertainty. However, the characteristics of CR and particularly of their
anisotropy are determined mainly by the {\em local} spatial and temporal distributions
of SNR. The actual distribution of 81 SNR, built in \cite{Case} is shown in 
Figure 6. It is more complicated than that described by the fit (1). In the vicinity of
 the Sun at $R_\odot > 8.5kpc$ there is an indication of a peak (~or even two~) which 
can be connected with the presence of the Orion arm, mentioned also in \cite{Case}. It 
is evident that the fit used in \cite{Case} is too simple and not able to reproduce the
 local flattening of the SNR radial distribution in the region of the Orion arm. 
\begin{figure}[htb!]
\begin{center}
\includegraphics[height=8cm,width=8cm]{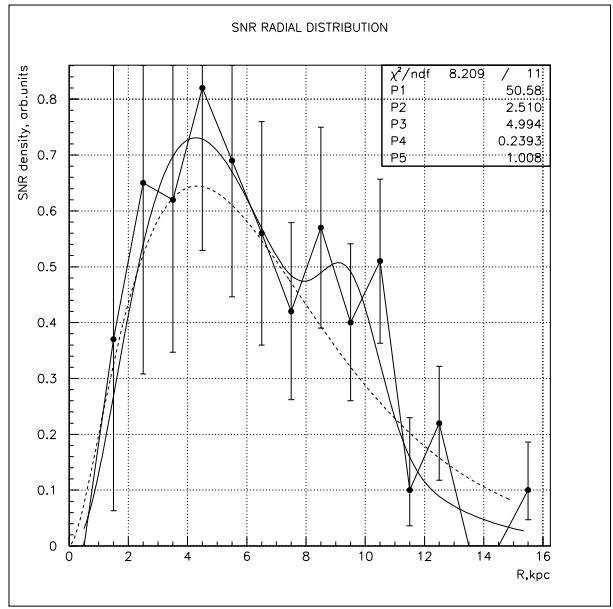}
\caption{\footnotesize Radial distribution of SNR and their fits. Saw-like curve -
the experimental distribution from \cite{Case}.  Errors are calculated for the 
Poissonian distribution applied for a small statistics. Dashed line - the fit of the 
distribution by the function (1). The full line - its fit by the 4-parameter function 
(9).}
\end{center}
\label{fig:anis6}
\end{figure}    

Since this local flattening can be important for the phase of the anisotropy we suggest
 using a more complicated 4-parameter fit:
\begin{equation}
\rho_{SN} \propto (\frac{R}{R_\odot})^a exp(-b\frac{R}{R_\odot})+c \cdot exp(-0.5(\frac{R-9500pc}{d})^2)
\end{equation}
It is shown in Figure 6 by the full line. The best fit parameters are: 
$a=2.5103 \pm 0.5645; b=4.9935 \pm 0.8376; c=(4.7314 \pm 2.8570)\cdot 10^{-3}; 
d=1008 \pm 439 pc$. They have been obtained using the Minuit code incorporated in the 
CERN program PAW. The poor accuracy of the parameters is due to the poor statistics 
of the experimental SNR distribution: only 81 SNR over 16 points.  Nevertheless, the 
quality of the 4-parameter fit: $\chi^2 = 0.7462$ is better than for the fit (1) used 
in \cite{Case}: $\chi^2 = 0.9084$. Certainly we cannot claim that that the original fit
(1) and our fit (9) are inconsistent with each other, but there are some advantages 
of our modified fit. The first one is that if the bump at the galactocentric 
distances $\sim$8 - 10 kpc is associated with the Orion arm then the Sun is localized 
at the inner edge of it, which corresponds to astronomical observations and gives a 
better description of the Sun's environment. Another one is that the use of our fit 
allowed us to achieve, in some cases, the correct phase of the anisotropy, which has 
not been obtained with the fit (1). It will be shown below. 

Some further remarks can be made about the spatial distribution of SNR. Concerning the 
large scale variation, with its fall - with increasing $R$ - from a peak surface 
density at $R\sim 4$ kpc, there can be no doubt. All the astrophysically related 
quantities (~pulsars, OB associations, molecular hydrogen etc.~) show at least the fall
 \cite{Fatem}. Turning to the important peak in the Outer Galaxy at $R\sim 10$ kpc 
(~10.5 kpc for the saw-like curve in Figure 6 and $\sim$9 kpc for the full line fit in 
the same Figure~) it is to be identified with the next spiral arm outwards. This 
feature is particularly noticable in the second quadrant (~$l: 90^\circ - 180^\circ$~) 
for many tracers of stellar activity.
 
We repeated the simulations of the anisotropy described above with the new radial 
distribution function (9). The result is shown in Figure 7. 
\begin{figure}[htb!]
\begin{center}
\includegraphics[height=10cm,width=15cm]{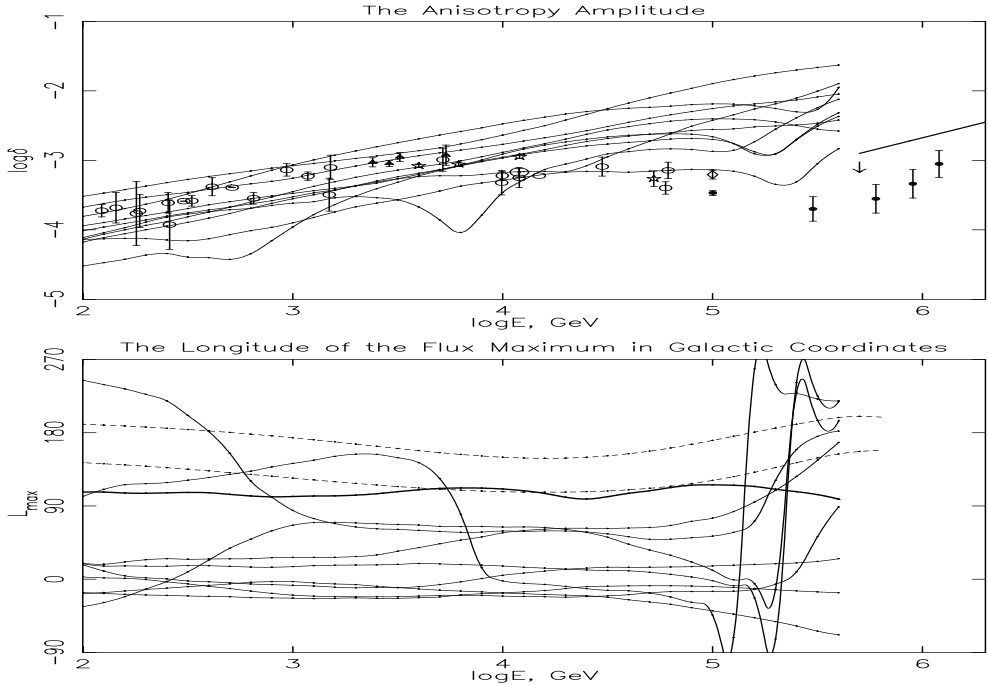}
\caption{\footnotesize The amplitude and the galactic phase of the CR anisotropy 
similar to those in Figure 5a and 5c, but simulated with the SNR radial distribution
(9) which takes into account the vicinity of the Orion arm. (a) The anisotropy 
amplitude: symbols for the experimental points are explained in the caption of Figure 
5. Thin lines are anisotropy amplitudes for 10 different configurations of SNR 
including a Single Source and the effect of the Halo. (b) The galactic phase of the 
anisotropy for the same 10 simulated samples of SN as shown in (a). Dashed lines 
denote the region of galactic longitudes which corresponds to the RA region populated 
by the experimental points shown in Figure 5b, assuming that the observations were made
 at declinations between 30$^\circ$ and 60$^\circ$. The thick full line indicates the 
only sample which has a phase in the second quadrant up to the highest energy of 
$4\cdot 10^5$GeV.}
\end{center}
\label{fig:anis7}
\end{figure}    

It is seen that due to the reduction of the CR gradient caused by the presense of the 
Orion arm the anisotropy amplitudes decrease in the mean and some of them approach the 
low values observed at energies between $10^4$ and $10^5$GeV. Due to the same reduction
 of the local gradient the role of the random SNR explosions grows and fluctuations 
of the anisotopy phases increase. Remarkable, among 10 simulated samples there 
is one which has the maximum flux of CR streaming from the Outer Galaxy, in particular 
from its second quadrant in the wide energy region from 10$^2$ up to 10$^5$ GeV 
(~thick full line in Figure 7b~). There 
was {\bf no} such samples in any of the previous simulations where CR fluxes either 
flow only from the Inner Galaxy or switch to the Outer Galaxy only in the limited 
energy range no wider than 1 - 1.5 decades (~Figure 5c~). 

\section{Discussion}

As has been shown in previous sections we are close to understanding the observed  
characteristics of the CR anisotropy below the knee in terms of CR origin from
SNR. The major features of the anisotropy are mainly determined by the local 
environment and its recent history in the vicinity of the solar system, say, within 
1 - 2 kpc from the Sun. That is the major factor. Other factors which have been 
examined here and turned to be helpful for reducing the anisotropy are the mixed mass 
composition and isotropic CR from the Galactic Halo. 

The anomalous diffusion in the non-uniform ISM adopted here magnifies the
role of the nearby sources. The CR flux from the Single Source if it is similar to the 
Monogem Ring SNR which is located in the Outer Galaxy compensates partly the flux from 
the bulk of SNR in the Inner Galaxy and reduces further the amplitude of the anisotropy
 at energies approaching 10$^5$GeV. 

It is in principle possible that these factors together with the accidental 
localization of SNR in space around the Sun and time of explosion are enough to ensure
the small and nearly energy independent amplitude of the anisotropy and its phase 
pointing out to the direction from the Outer Galaxy. However, the probability of such 
an accidental localization is low. In the process of our study we also simulated a 
number of SNR patterns with the real positions of SNR nearby the Sun. We used 
coordinates of SNR {\em observed} within 3 kpc from the Sun from the catalog of 
galactic SNR \cite{Green}. The number of SNR from the Inner and Outer half-circles in 
this pattern was approximately equal. We could not use the actual ages of 
these SNR since all they were quite young, viz. less than $4\cdot 10^4$ years, and 
according to our model of the SNR explosion \cite{EW1} CR are still confined inside 
their expanding shells. However, insofar as many - perhaps most - SN are produced in 
clusters, these SN can be used as potential sites for previous SN which are no longer 
visible. Thus, we attributed to each of them 100 random ages 
distributed uniformly within $10^7$ years in the past. Outside the 3kpc circle the SNR 
were simulated as in the main run, i.e. with the spatial distributions (1) and (2). 
In {\em none} of the patterns did we have an anisotropy phase from the Outer Galaxy.
 
In our mind it is the position of the solar system on the inner edge of the Orion 
spiral arm (~or 'sub-arm' as it might be termed, insofar as it is rather weak feature~)
 with its inverse local gradient of the star density, which helps to reverse the 
main flux of CR from the Inner Galaxy for the favorable pattern of positions and 
ages of nearby SNR. We understand that our model of the Orion arm is just a simplistic 
imitation and the real structure is much more complicated. Also the fact that only one 
 among ten simulated samples has the phase from the second quadrant of the Outer Galaxy
 is not an impressive agreement. However, taking into account on the one hand the 
uncertainty of the experimental data caused by the small magnitude of the anisotropy 
and on the other the absence of any information about the actual position and ages of 
SNR in our model which are not so young as those in the catalog \cite{Green}, we 
consider our model of the SNR radial distribution with the Orion arm (9) as a promising
 approach.

It is true that 10 samples shown in Figure 7 are not enough to give a precise estimate
of the probability for an accidental turn of the phase to the second quadrant. The 
low statistics are not only due to the large required computer time ($\sim$40 hours 
for 10 samples of just one primary nucleus~). In fact our work does not pretend to give
 a the precise quantitative 
description of the phenomenon, since the experimental data themselves spread over a 
large range and our treatment of the processes in the SNR and ISM are simplified. That 
is why we believe that in such a complicated phenomenon as the anisotropy even 
semi-quantitative estimates and indications are of value.

Saying this we admit that our solution seems to be not sufficiently radical.
Even if we assume that the distribution (9) is axisymmetric the fraction of SNR 
determined by the second term of it and associated with the Orion arm is only 15\%.
The $\sim10$\% probability of the successful description of the experimental data seems
 to be also too small. This feeling is supported by the estimates
of the gamma-ray gradient in the Galaxy \cite{EW8,EW9}. The ratio between the gamma-ray
intensities at galactic longitudes $\ell = 0^\circ$ and $180^\circ$ calculated with
the SNR radial distribution (1) for the energy above 0.1 GeV is about 10.9 while the 
experimental data of the EGRET collaboration indicate that it does not exceed 3 
\cite{Hart}. The introduction of the increased SNR density in the Orion arm (~actually
in the model it is 'the Orion circle' because the distribution (9) is axisymmetric~) 
reduces this ratio down to 9, but not to 3. This shows that the effect of 'our Orion
arm' is not big enough to reduce the {\em total} gradient, to which the gamma-ray 
intensity is more sensitive than the anisotropy which is more sensitive to the 
{\em local} CR gradient.

There are a number ways of reducing the total CR gradient which will also help to 
increase the probability of inverting the local gradient. The most radical way is to 
abandon the hypothesis about SN being the main source of CR and to find another source
with a flatter radial distribution. There are several such proposals, but we shall
 not discuss them here, relying on the more conservative hypothesis of SN origin. 

Another way is to assume the flatter radial distribution of SNR, than that given by 
Case and Bhattacharya \cite{Case} which might make it steeper due to an inefficiency of 
the SNR observation at large distances from the Sun. Some indication of such 
inefficiency can be seen in their map of 159 SNR (~Figure 1 in \cite{Case}~), where the
 SNR density decreased with the distance from the Sun not only toward the Outer Galaxy 
but also toward the first and fourth quadrant of the Inner Galaxy. 

One more way is to introduce the dependence of the diffusion characteristics on the 
galactic radius due to the different 'turbulisation' of the ISM by SNR explosions, 
which are more intensive in the Inner Galaxy \cite{EW10}.

Another proposal is to introduce a radial dependence of the acceleration efficiency.
Purely phenomenolically it means that the radial distribution of CR, $\rho_{CR}(R)$,
 should be connected with that of SNR, $\rho_{SN}$, by means of introducing an 
efficiency index
$\varepsilon < 1$ as $\rho_{CR} \propto \rho_{SN}^\varepsilon$. The estimates indicate
that for $\varepsilon = 0.5$ the ratio of CR and SN gradients in the vicinity of the 
Sun is 0.315, for $\varepsilon = 0.25$ it decreases down to 0.075. The observed CR 
gradient $\frac{d(ln\rho_{CR})}{dR}=-0.04\pm 0.03$ \cite{EW10} corresponds to 
$\varepsilon = 0.52^{+0.18}_{-0.26}$.  

There are models which use the phenomenon of the Galactic Wind, which is more intensive
in the regions with the higher rate of SN explosions and sweeps out more CR particles 
from the Galactic Disk there \cite{Breit,Voelk}. As a result, the radial distribution 
of CR in the disk becomes flatter and the gradient smaller. This class of models has
not the only advantage that it utilizes the well known phenomenon of galactic winds, 
but also because such winds initiate the formation of the Galactic Halo, which in turn
assists reducing the irregularity of the CR energy spectrum and the CR gradient 
\cite{Voelk,EW7}.         
  
If any of the proposed mechanisms is really working the probability of getting a
favorable local SN pattern, which ensures the needed small amplitude and phase of the 
CR anisotropy, can be more than 'our' $\sim 10$\%. We should also say that the 
observed anisotropy is a very complicated phenomenon. At the same time {\em different}
 SNR can contribute {\em different} nuclei at {\em different} energies. In order to 
unveil the specific spatial and temporal distribution of these SNR or other local 
objects which determine the observed characteristics of CR one should study the energy 
dependence of the anisotropy amplitude and phase for separate primary nuclei. This is
a practical possibility for future experiments. 

\section{Conclusions}

We have shown that if the bulk of the Cosmic Radiation is produced by shocks from SN 
explosions, the observed 
characteristics of their anisotropy are the product of local phenomena. These 
explosions are a stochastic process and the observed characteristics appear to some 
extent as accidental. The small magnitude and the nearly energy 
independent amplitude of the anisotropy can be the consequence of the mixed mass 
composition, the influence of the Single Source - the nearby and recent SN explosion, 
the smoothing effect of the Galactic Halo and the position of the Sun on the 
inner edge of the Orion arm. In order to unveil the specific spatial and temporal 
distribution of nearby SNR and other local objects which determine the 
characteristics of the observed CR one should study the energy dependence of the 
anisotropy amplitude and phase for separate primary nuclear charges.         
    
{\bf Acknowledgments}

The Royal Society and the University of Durham are thanked for financial support.

\end{document}